\documentclass[12pt]{iopart}
\usepackage{epsfig}
\usepackage{amssymb}

\def\la{~\mbox{\raisebox{-.6ex}{$\stackrel{<}{\sim}$}}~}
\def\ga{~\mbox{\raisebox{-.6ex}{$\stackrel{>}{\sim}$}}~}
\newcommand{\be}{\begin{eqnarray}}

\newcommand{\ee}{\end{eqnarray} }

\begin{document}

\rightline{UMN--TH--3202/13}

\title{A review of Axion Inflation in the era of Planck}

\author{Enrico Pajer$^1$ and Marco Peloso$^{2,3}$}

\address{
$^1$ Department of Physics, Princeton University, Princeton, NJ 08544, USA
}

\address{
$^2$ School of Physics and Astronomy, University of Minnesota, Minneapolis, MN 55455, USA
}

\address{
$^3$ INFN, Sezione di Padova, via Marzolo 8, I-35131, Padova, Italy
}


\begin{abstract}
Because the inflationary mechanism is extremely sensitive to UV-physics, the construction of theoretically robust models of inflation provides a unique window on Planck-scale physics. We review efforts to use an axion with a shift symmetry to ensure a prolonged slow-roll background evolution. The symmetry dictates which operators are allowed, and these in turn determine the observational predictions of this class of models, which include observable gravitational waves (potentially chiral), oscillations in all primordial correlators, specific deviations from scale invariance and Gaussianity and primordial black holes. We discuss the constraints on this class of models in light of the recent Planck results and comment on future perspectives. The shift symmetry is very useful in models of large-field inflation, which typically have monomial potentials, but it cannot explain why two or more terms in the potential are fine-tuned against each other, as needed for typical models of small-field inflation. Therefore some additional symmetries or fine-tuning will be needed if forthcoming experiments will constrain the tensor-to-scalar ratio to be $r \la 0.01$.
\end{abstract}


\maketitle

\section{Introduction}
\label{sec:intro}

The recent results from the Planck satellite \cite{PlanckI} are remarkably compatible with the most minimal single field slow-roll inflationary models. There is a simple argument that shows that inflationary models that lead to enough efoldings to be viable, namely around 50 to 60, are very sensitive to Planck-scale physics\footnote{In this discussion by ``Planck scale'' we mean the scale at which our effective four-dimensional description of gravity plus one or more scalar fields breaks down. Depending on the models under consideration this could be some Kaluza-Klein scale, the string scale or something even more exotic. The main point is that there is some scale of new physics and it has to be at or below the (reduced) Planck mass $M_{Pl} \equiv \left( 8\pi G_{N} \right)^{-1/2}$.  If this scale is below $M_{Pl}$,  our arguments about UV-sensitivity become even stronger.}. Imagine writing down some potential that induces 60 efoldings of expansion, call it $V_{sr}$ where ``$sr$'' stands for slow roll. Then consider all dimension five and six operators involving the inflaton. Unless forbidden by a symmetry, these operators, such as for example $\phi^{2} V_{sr}/M_{Pl}^{2}$ induce corrections of order $\mathcal{O}(1)$ or larger to the slow-roll parameter $\eta$ hence drastically shortening the duration of inflation to just a few efoldings. This means that one needs to ensure that whatever physics there is above the Planck scale, it does not induce these terms (or it largely suppresses them). This \textit{UV-sensitivity} is present for any inflationary potential, but becomes much more dramatic for models, which we will call large field models, in which the inflaton vacuum expectation value (vev) changes by an amount much larger than the Planck scale (or whatever the scale of new physics is) during inflation. In that case Planck suppressed operators of any dimension (e.g.$\phi^{n}M_{Pl}^{4-n}$) can dramatically change the dynamics of inflation. This means that one needs to control an infinite number of higher dimension operators. In summary, in order to claim that a given model of inflation is theoretically robust one needs to understand why potentially dangerous corrections coming from heavy modes above the Planck scale are suppressed or forbidden.

It has long been recognized (see e.g.~\cite{CLLSW,ACCR2} for two clear discussions) that supersymmetry is not sufficient to protect slow-roll inflation from radiative corrections because it is broken by the inflationary background at the Hubble scale. Although it is an interesting question whether the mechanism that ensures the flatness of the potential, e.g. a shift symmetry to be discussed shortly, can be embedded in a SUSY model, we will refrain from discussing the details of supersymmetric constructions here. Instead we refer the reader to the extensive literature \cite{othermodels}.

The only symmetry that can forbid the sort of corrections discussed above is a \textit{shift symmetry}, i.e.~one assumes that the action is invariant under a transformation like $\phi\rightarrow \phi+$const. We will refer to a field possessing this symmetry (at least to some approximate level) as an \textit{axion}. The first model of axion inflation was proposed a long time ago and named \textit{natural inflation} \cite{natural}. Since then several other models have been proposed and studied exploiting the protection of the shift-symmetry. We will discuss these models in section \ref{sec:models}. One scale that plays an important role in all axion models is the \textit{axion decay constant} $f$. This can be though of as determining the strength of the least irrelevant shift-symmetric coupling, such as the dimension five coupling to gauge fields taking the form $\phi F \tilde F/f$. Instantons in the gauge sector (and their string theory cousins) break the continuous shift symmetry down to a discrete one $\phi\rightarrow \phi+2\pi f$ at the non-perturbative level. In this sense $f$ can be though of as the periodicity (up to $2\pi$) of the axion potential in the absence of any explicit breaking of the shift-symmetry. 

Therefore, predictions for axion inflation are under theoretical control for
\begin{equation}
m_\phi , H < f < M_{Pl}
\label{window-f}
\end{equation}
with $H$ being the Hubble parameter and $m_{\phi}$ the mass of the axion. The lower limits are obtained from the fact that, in general, the theory of the axion arises by integrating out modes that are heavier than $f$, and therefore it
can describe only dynamics at lower scales.  The upper bound comes from  various considerations. One is that quantum gravity effects are expected to break the shift symmetry, as any global symmetry, at the scale $M_{Pl}$ \cite{Planckcorr} throught the formation and subsequent evaporation of a (virtual) black hole\footnote{Potentially some additional suppression can appear because of the non-perturbative nature of the effect  \cite{KKLS}.}. This argument does not apply if the shift symmetry emerges from a gauge symmetry as typically in string theory. However, also in this case the limit appears to apply, since all known  controlled string theory constructions are characterized by   $f<M_{Pl}$ \cite{BDFG,SW}. There are many different ways to get four-dimensional axions by dimensional reduction of some compactified string theory, but in all cases $f/M_{Pl}$ turns out to be proportional to positive powers of some small control parameter and hence cannot be larger than $1$. A simple example are Type IIB axions coming from integrating an RR two-form over a two cycle. In this case one finds $f/M_{PL}\simeq \sqrt{g_{s}}/L^{2}$ where $g_{s}$ is the string coupling and $L$ is the length scale of the compact two cycle defining the axion in string units. Notice that these constructions are understood only at weak coupling $g_{s}\ll1$ and for geometric compactifications in which $L\gg1$, leading to $f/M_{Pl}\ll1$. Discussions on how these bounds could have a more fundamental origin are given in \cite{V}.

\begin{figure}
\centering
\includegraphics[width=\textwidth]{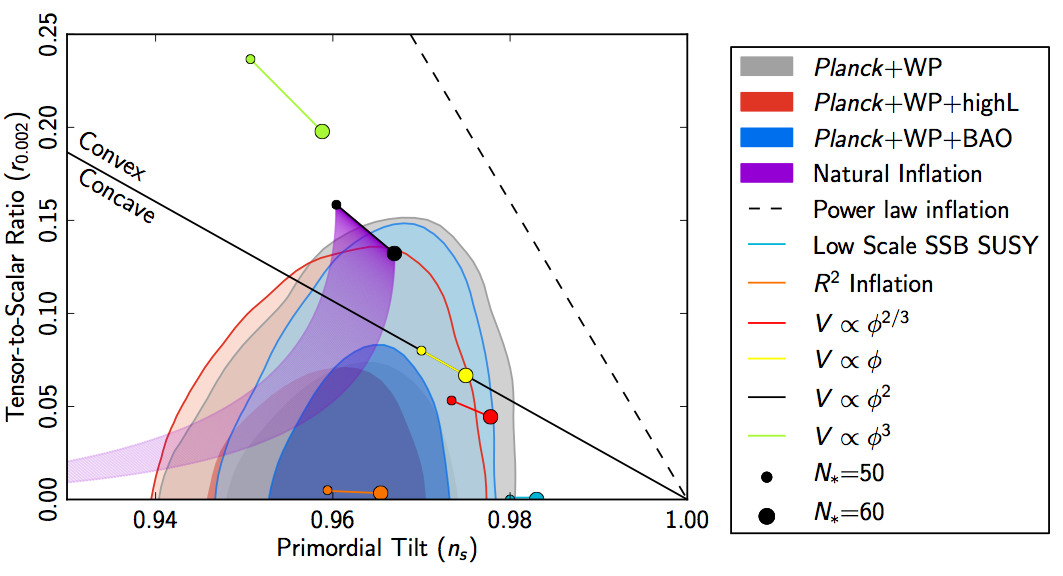}
\caption{The figure, taken from \cite{PlanckInfl} shows the $68\%$ and $95\%$ CL contours form Planck plus various ancillary sets of data (indicated in the right panel) in the tensor-to-scalar ratio $r$ vs scalar spectral tilt $n_{s}$. Shown are also several inflationary models. In the main text we discuss those potential that arise for axions, i.e.~$\phi^{2/3}, \phi,\phi^{2}$ and natural inflation.\label{fig1}}
\end{figure}

Models of axion inflation lead to several characteristic signals in late time observables as we will discuss in section \ref{sec:pheno}. Two of the most basic observables are the tilt of the scalar power spectrum $n_{s}$ and the tensor-to-scalar ratio $r$. Figure \ref{fig1}, borrowed from \cite{PlanckInfl}, shows the constraints in the $r-n_{s}$ plane from the CMB spectrum as measured by Planck in combination with CMB polarization from WMAP and BAO or high $l$ measurements from ACT or SPT. The marginalized upper bound on $r$ implies that primordial gravitational waves have less than $11\%$ of the amplitude of scalar perturbations \cite{PlanckInfl}, but several balloon borne and ground based experiments will improve this bound by at least an order of magnitude or make a detection in the near future   \cite{Baumann:2008aq}.  The marginalized bounds on the spectral tilt from Planck plus WMAP are $0.9457<n_{s}<0.9749$ at $95\%$ CL \cite{PlanckInfl}.

As we discuss  in subsection \ref{ss:natural}, the simplest implementation of axion inflation (namely,  natural inflation \cite{natural}) results in a too red spectrum ($n_s$ too small) if the   theoretical bound $f<M_{Pl}$ is respected. A number of models have therefore been proposed that effectively result in a superPlanckian displacement for the axion even if  $f<M_{Pl}$ (small-field models with subPlanckian displacement are discussed in section \ref{sec:discussion}). These proposals have the additional advantage that they typically can result in an observable gravity wave signal. In fact, the excursion of the inflaton in field space is related to the amount of generated gravity waves from vacuum fluctuations by the Lyth bound  \cite{L} for canonical fields (for a more general case see \cite{BG}): 
\be
\frac{\Delta \phi}{M_{Pl}}>\frac{N_{CMB}}{30} \sqrt{\frac{r}{0.009}}
\ee
which shows that the observationally relevant region  cannot be achieved for $\Delta \phi \ll M_{Pl}$. We will also discuss models in which tensors are sourced directly and hence their amplitude is not subject to the Lyth bound,  as happens e.g.~in presence of the shift-symmetric coupling $\phi F \tilde F$.

We review several  of these models in section \ref{sec:models}. Many of these models present    several interesting potentially observable signatures, such has  oscillations in all primordial correlators, violation of scale invariance, non-Gaussianity, primordial black holes and chiral gravitational waves. A crucial point worth stressing is that all these signals arise because of the underlying shift symmetry. Because models of axion inflation rely on the shift symmetry to ensure a flat potential, all these observables are a direct consequence of having a theoretically robust inflationary model. Also, these observables are \textit{correlated} in the sense that they all arise in the \textit{same} theory, i.e.~the theory of one or more axions. The search for the signals predicted in axion inflation has received a lot of attention in the literature and we will discuss in detail these efforts in section \ref{sec:pheno}. Finally section \ref{sec:discussion} contains a discussion of the role of the shift symmetry in small-field models.


\section{Models}
\label{sec:models}

In this section we review the construction of some of the most popular models of axion inflation. We give a summary of the main ingredients and critically discuss their shortcomings. We present most of the results without a derivation, referring the reader to the original literature for details. We first discuss the original model of axion inflation, namely natural inflation \cite{natural}. As we mentioned, this model is compatible with phenomenology only for $f \gg M_{Pl}$. 

Several  directions have been followed to reconcile this with the theoretical  requirement (\ref{window-f})  of a subPlanckian axion decay constant.  One  is to break the shift symmetry in a controlled way, either explicitly or spontaneously, as we discuss  in subsections \ref{ss:monodromy} and \ref{ss:4}; another one is to use more than one axion, as we discuss in subsection \ref{ss:many}; another one is to invoke some additional dynamics that arises from the coupling to other fields (in particular, gauge fields) as in subsection \ref{ss:gauge}; one final possibility is to use non-local operators that arise in extra dimensional contexts, as we discuss in subsection  \ref{ss:extradim}.

 
\subsection{Natural Inflation}
\label{ss:natural}

In the absence of any explicit breaking of the shift-symmetry, the potential for the axion is exactly constant at all orders in perturbation theory. In order to provide an inflationary potential, the authors of \cite{natural} (see also \cite{ABFFO}) have proposed Natural inflation that relies on the non-perturbative breaking of the axion shift-symmetry by gauge theory instantons. The resulting potential takes the form
\be\label{natpot}
V=\Lambda^{4} \left[1+\cos \left(\frac{\phi}{f}\right)\right]\,,
\ee
where $\Lambda$ is some non-perturbatively generated scale, namely proportional to $e^{-1/g}$ for some gauge coupling $g$ and $f$ is the axion decay constant. Notice that $\phi$ here and in the rest of this review has always a canonical kinetic term. In principle higher instantons induce higher harmonics corrections to this potential, but with an amplitude that is suppressed with respect to the leading term by powers of $\Lambda/M_{Pl}$, with $M_{Pl}$ a proxy for some UV-cutoff scale. We will always assume $\Lambda \ll M_{Pl}$ and safely neglect these small corrections in the rest of the paper.

Figure \ref{fig1} shows how natural inflation compares against the Planck results.  Natural inflation makes a one parameter family of predictions depending on the value of the axion decay constant $f$. Comparison with WMAP3 (see \cite{ABFFO,MT} for earlier bounds) with flat priors on $f$ and fixing $N=60$ lead to $f>3.7 \times M_{Pl}$ at $95\%$ CL \cite{FKS}~\footnote{We note that the limit in \cite{FKS} is expressed in terms of the  non-reduced Planck mass which is $\sqrt{8\pi}\sim 5$ times larger than $M_{Pl}$. Also notice that the same analysis, i.e.~fixed $N=60$ and a flat prior on $f$, gives $f<4.3 \times M_{Pl}$ using Planck data.}. With Planck this gets updated to $f> 10  \times M_{Pl}$ at 95 \% CL \cite{PlanckInfl} for different priors that are discussed in details in \cite{EMP,PlanckInfl}. The tension between these observational requirements and the theoretical bound (\ref{window-f}) motivates the proposals that we discuss in the reminder of this section.

 
\subsection{Axion monodromy inflation}
\label{ss:monodromy}

In \cite{SiW} it was noticed that by breaking the shift symmetry explicitly the field space opens up and arbitrary large excursion of the axion are allowed. This is tantamount to assuming the presence of some \textit{monodromy}, hence the name axion monodromy inflation. If the breaking can be made small in an appropriate sense, additional corrections to the axion potential can be neglected and the leading shift-symmetry-breaking term drives inflation.

The first realization of this idea involved D-branes moving around a Nil manifold in type IIA string theory and, after imposing additional symmetries to ensure the flatness of the potential, produced a potential $\phi^{2/3}$, whose predictions are shown in figure \ref{fig1}. A different construction using model-dependent (i.e.~coming from integrating $p$-forms on compact $p$-cycles) axions in type IIB was proposed in \cite{MSW} and further studied in \cite{FMPWX,Dante} (see also \cite{C}). The candidate axion is obtained integrating the RR two-form\footnote{One additional difficulty in embedding the monodromy effective potential in string theory is that the stabilization of the moduli appearing after dimensional reduction makes use of non-perturbative effects that lead to additional breaking of the shift symmetry. The non-perturbative effects create a potential for the NSNS two-form that make it an unsuitable inflaton candidate. \cite{MSW}} over a two-cycle. The shift symmetry is broken by some NS5-brane wrapping the same two-cycle (with a tadpole-canceling anti-NS5-brane wrapping a distant homologous two-cycle). In order to maintain control over the construction one needs the breaking of the shift symmetry and of supersymmetry, both induced by the NS5-brane, to be small. To achieve this the two-cycle in question needs to sit in a warped region. It should be noted that explicit constructions of the ensuing backreacted geometry are not known explicitly. Finally, the non-perturbative effects used in moduli stabilization lead to additional corrections to the inflaton potential (remember that the shift-symmetry protects only against perturbative corrections) that can be controlled by imposing some isometry on the geometry of the throat. A different resolution of this problem that makes the potential more robust against any other non-perturbative correction involves a second axion and leads to a potential reminiscent of Dante's Inferno, as we will see in section \ref{ss:many}.

The resulting potential comes from the DBI action of the NS5-brane and takes the form
\be\label{amp}
V(\phi)=\mu^{3} \left( \sqrt{\phi^{2}+\phi_{c}^{2}}-\phi_{c}\right)\,,
\ee
where  $\mu,\phi_{c}\ll M_{Pl}$ are given explicitly in section VI.A of \cite{BPP}. Notice that for $\phi\gg \phi_{c}$ the potential becomes linear and its predictions are shown in figure \ref{fig1}. In \cite{DHSW} it was argued that  potentials that are less convex than quadratic, namely of the form $\phi^{p}$ with $p<2$ as in Eq.(\ref{amp}), are generically produced in this context through interactions with heavy fields. The application of the axion monodromy potential in Eq.(\ref{amp}) to quintessence was discussed in \cite{PST}. Finally it should be noticed that non-perturbative corrections that are independent from moduli stabilization can lead to an additional sinusoidal term in Eq.(\ref{amp}) of the form of Eq.(\ref{natpot}). In the string theory construction these arise e.g.~from ED1 instanton corrections to the K\"ahler potential. The size of the oscillating correction is non-perturbatively small and hence it does not pose a threat to the inflationary dynamics. Instead it leads to a very unique signal in all primordial correlators to be discussed in subsection \ref{ss:res}.

 
\subsection{Axions coupled to a 4-form}
\label{ss:4}

An interesting possibility is to break the shift symmetry spontaneously. A concrete realization of this idea was considered in \cite{KS,KLS} (see also \cite{KS2} for applications to quintessence). The starting point it to consider the coupling of the axion to a four form
\be
S=\int d^{4}x \sqrt{g} \left[\frac{M_{Pl}^{2}}{2}R-\frac{1}{2}\partial_{\mu}\phi\partial^{\mu}\phi - \frac{1}{48}F_{\mu\nu\rho\sigma}F^{\mu\nu\rho\sigma}+\frac{\mu}{24}\phi \frac{\epsilon_{\mu\nu\rho\sigma}F^{\mu\nu\rho\sigma}}{\sqrt{g}}\right]\,,
\ee
where $F_{\mu\nu\rho\sigma}=\partial_{\mu} A_{\nu\rho\sigma}$ (equivalently $F_{4}=dA_{3}$). Upon integration by part one can see that the new coupling respects the shift symmetry at the level of the action and that a vev for $F_{4}$ will break it spontaneously. It is convenient to rewrite the action in terms of the dual magnetic variable to $F_{4}$, which is a scalar that we will call $q$. To do this we can follow similar steps as in the Palatini formulation of gravity. We assume that $F_{4}$ and $A_{3}$ are two distinct fields in the action and add a Lagrange multiplier to impose $F_{4}=dA_{3}$. We can then integrate out $F_{4}$ since it now appears quadratically (after completing the square). The final action takes the form
\be
S=\int d^{4}x \sqrt{g} \left[\frac{M_{Pl}^{2}}{2}R-\frac{1}{2}\partial_{\mu}\phi\partial^{\mu}\phi - \frac{1}{2}(q+\mu\phi)^{2}+\frac{1}{6} \frac{\epsilon^{\mu\nu\rho\sigma}A^{\nu\rho\sigma}\partial_{\mu}q}{\sqrt{g}}\right]\,.
\ee
Notice that now the shift-symmetry is realized by shifting both $q$ and $\phi$ at the same time and it is broken when $q$ gets a vev. In this case the potential is just a shifted $m^{2}\phi^{2}$ with the $\mu$ playing the role of the mass. The shift symmetry ensures that the mass is protected against quantum corrections and other corrections from higher dimensional operators have been argued to be small. Non-perturbative corrections can again induce a periodic modulation on top of this potential to be discussed in subsection \ref{ss:res}.

 
\subsection{Inflation with more than one axion}
\label{ss:many}

Instead of breaking the shift-symmetry, a superPlanckian excursion of the inflaton can be achieved if inflation is driven by more than one axion. The first implementation of this idea was given in \cite{KNP} where it was assumed that two axions are present and they interact through some non-perturbative potential
\be
V(\rho,\theta)=\Lambda^{4}_{1} \left[1-\cos \left(\frac{\rho}{f_1}+\frac{\theta}{g_{1}}\right)\right]+\Lambda^{4}_{2} \left[1-\cos \left(\frac{\rho}{f_2}+\frac{\theta}{g_{2}}\right)\right]\,,
\ee
where $f_{1,2}$ and $g_{1,2}$ are different axion decay constants. For $f_1 / f_2 = g_1 / g_2$ the same linear combination of the axions enters in both terms, and the orthogonal combination is massless. When this relation holds approximatively, one combination is substantially lighter than the ``naive expectation'' $\Lambda_i^2 / f_{i} , \Lambda_{i}^2 / g_{i}$. Inflation can be driven by the lighter combination, with the heavier one 
settled to a minimum.  For simplicity, let us assume $f_1=f_2 =f$. The potential along the light direction takes the form
\be
V(\xi)=\Lambda^{4}_{2} \left[1-\cos \left({\rm const}+\frac{\xi}{ f_{\xi}}\right)\right]\,,
\ee
where $\xi\propto (\frac{\rho}{f})-(\frac{\theta}{g_{1}})$ and the effective axion decay constant is $f_{\xi}=g_{2}\sqrt{f^{2}+g_{1}^{2}}/|g_{1}-g_{2}|$. Allowing for some tuning of $g_{1}$ and $g_{2}$ one can make $f_{\xi}$ arbitrarily large, hence achieving an effective superPlanckian axion decay constant and the phenomenological predictions of natural inflation shown in figure \ref{fig1}.

A different model of inflation with two axions called Dante's inferno was constructed in \cite{Dante} as a generalization of the string theory construction discussed in subsection \ref{ss:monodromy}. One assumes that the ED1 instanton that induces periodic modulations of the axion monodromy potential, extends over some linear combination of the cycle wrapped by the NS5-brane and another cycle. Only the axion $\rho$ associated with the former cycle posses a monodromy, while the latter, $\theta$, enjoys a discrete shift-symmetry 
\be
V(\rho,\theta)=W(\rho)+\Lambda^{4} \left[1-\cos \left(\frac{\rho}{f_{\rho}}-\frac{\theta}{f_{\theta}}\right)\right]\,,
\ee
where $f,g\ll M_{Pl}$ are the two axion decay constants and $W(\rho)$ is the monodromy potential. As we mentioned before it is not always easy to keep $W$ slow-roll flat because of the use of non-perturbative effects in the stabilization of moduli. Nevertheless, even when $W$ does not satisfy the slow-roll conditions, the above potential can induce a prolonged stage of inflation under two conditions. First, $V$ in the $\rho$ direction must be non-monotonic, which happens when the modulation proportional to $\Lambda$ stops the descent caused by $W$, which can be easily satisfied in the string theory model. Second one needs a hierarchy between the two axion decay constants, namely $f_{\rho}\ll f_{\theta}$. In this regime the dynamics proceeds along a linear combination $\phi=\theta+\rho f_{\rho}$, which is parametrically flatter than the $\rho$ direction. The algebraically simplest case (although this mechanism works for more general power laws) is a quadratic potential $W(\rho)= m^{2} \rho^{2}/2$, in which case the effective potential for $\phi$ is again quadratic with an effective mass $m_{\rm eff}\equiv m f_{\rho}/f_{\theta}\ll m$. 
A remarkable fact about this dynamics is that, although the length of the inflationary trajectory is superPlanckian, say $\Delta \phi \sim 15\times M_{Pl}$ for a quadratic potential, the trajectory is contained in a region of field space that is subPlanckian in diameter, when written in terms of the fundamental fields $\rho$ and $\theta$ on which the symmetries act in a simple way \cite{Dante}. In fact, for a mild hierarchy $f_{\rho}/f_{\theta}\sim 10^{-2}$ one finds $\Delta \theta=2\pi f_{\theta}$ and $\Delta \rho \simeq \Delta \phi f_{\rho}/f_{\theta} < M_{Pl}\ll \Delta \phi$ and hence corrections to $W(\rho)$ away from a quadratic monomial, which is the generic leading term around a minimum, can be neglected. This implies that even producing observable tensor modes and satisfying the Lyth bound, this two field model is unaffected by the theoretical difficulties associated with large field excursion.

Let us now discuss the model with a large number of axions, called axion N-flation, which was proposed in \cite{N-flation} and further discussed in \cite{EM,G,KSS,O,BB,KLS2}. Ignoring higher instanton terms, the potential can be written as
\be
V(\phi_{n})=\sum_{n=1}^{N} \Lambda^{4}_{n} \cos \left(\frac{\phi_{n}}{f_{n}}\right)+{\rm const}\,,
\ee
again with canonically normalized fields. Notice that although each axion starts with a subPlanckian displacement from its minimum $\Delta \phi_{n} \equiv \phi_{n}-\phi_{n}^{(0)}\sim f_{n} \ll M_{Pl}$, the total displacement in the $N$-dimensional space is the diagonal of a hypercube. For example, if all $f_{N}$ were equal to $f$ the total displacement $|\Delta \phi_{n}|\sim \sqrt{N} f$ can easily be superPlanckian for large enough $N$. The predictions of the model depend on the distribution of $f_{n}$. When they are all equal the predictions for the tensor and scalar power spectrum are indistinguishable from $m^{2}\phi^{2}$ (see figure \ref{fig1}). A detailed analysis of a more general case was carried out in \cite{EM} using Random Matrix Theory (RMT) tools and information from realistic string theory constructions. The result is that if the axions constitute a sizable fraction of the total number of fields (moduli), the spectral tilt can be more red than in standard quadratic inflation, typically by some $25\%$. Several attempts were made to embed this model in string theory (e.g.~\cite{G,KSS,O,P}) but different obstacles such as backreaction make it very hard to control the large $N$ limit. One point worth stressing is that loop corrections induced by the $N$ fields renormalize the Planck scale \cite{N-flation} in such a way to prevent inflation for very large $N$. Because of this, there is only a model-dependent window of values of $N$ in which successful N-flation can take place. The possible production of non-Gaussianity was discussed in \cite{KLS2}.

Finally we should mention the models in \cite{ACCR2,KW,BG2} where a shift-symmetric field is coupled to a waterfall field to realize hybrid inflation. One difficulty consists in preventing the coupling between the two fields from generating large radiative corrections. Ideas from particle physics models such as supersymmetry and little Higgs are used to overcome this obstacle. It is important to notice that these are the only construction discussed here that lead to small-field inflation with unobservable tensor modes. We will come back to this in section \ref{sec:discussion}.

\subsection{Inflation from additional dynamics} 
\label{ss:gauge}

Although the potential of natural inflation is too steep for $f<M_{Pl}$, several authors have proposed to use additional friction to ensure a slow-roll evolution. A shift-symmetric coupling to gauge fields provide a simple and controllable realization of this idea and are discussed in the following. Shift-symmetry-violating couplings to a large number of fields have also been proposed \cite{trapped}. In this model, called trapped inflation, quanta of the extra fields are produced hence slowing down the inflaton. For a modification of gravity to implement an analogous mechanism see \cite{GK}. All these construction lead to small-field models of inflation, namely $\Delta \phi \sim f < M_{Pl}$. The shift symmetry is then not really necessary nor sufficient to ensure 60 efoldings (see discussion in section \ref{sec:discussion}) and additional symmetries or a fine tuning of about $1/100$ are needed to obtain a viable model. 

 
\subsubsection{Coupling to an Abelian gauge field}

Starting from ref.  \cite{Anber:2006xt}, it has been recently understood that the inflaton coupling to a gauge field through the shift-symmetric operator $\frac{\alpha}{ 4 f} \phi F {\tilde F}$ (where $\alpha$ is a dimensionless coefficient) can  play a nontrivial role already during inflation. In fact, due to this coupling,  one helicity of the gauge field experiences a tachyonic growth. The two gauge field polarizations satisfy \cite{Anber:2006xt}
\begin{equation}
\left( \frac{\partial^2}{\partial \tau^2} + k^2 \mp 2 a H k \xi \right) A_\pm \left( \tau , k \right)=0 \,,\quad \mathrm{with}\quad
\xi \equiv \frac{\alpha \dot{\phi}}{2 f H}
\label{eq-A}
\end{equation}
where $\tau$ is conformal time, $a$ the scale factor of the universe, $H$ the Hubble rate and $k$ the comoving momentum of the mode. For definiteness, we assume that $\xi > 0$, so that the mode that experiences the tachyonic growth is the $+$ one. As we shall see, $\xi \gg 1$ is phenomenologically excluded, while $\xi \la 1$ leads to unobservable effects. Therefore, $\xi \ga 1$ is considered here. In this case, the + polarization becomes tachyonic close to horizon crossing, and this takes place for modes of progressively large $k$ as inflation proceeds.  The physical energy density in each mode  rapidly grows at 
horizon crossing, but it  eventually gets diluted away by the expansion of the universe; therefore, at each moment during inflation, only modes with $k/a \sim H$ are present. Eq.(\ref{eq-A}), with $\xi$ constant, is solved by Coulomb functions. The solution   with initial conditions in the adiabatic vacuum is well approximated by  \cite{Anber:2006xt}
\begin{equation}
A_+ \left( \tau , k \right) \simeq \frac{1}{\sqrt{2 k}} \left( \frac{k}{2 \xi a H} \right)^{1/4} {\rm e}^{\pi \xi - 2 \sqrt{2 \xi k / \left( a H \right)}}
\;\;, \xi \ga 1
\label{A-sol}
\end{equation}
in the interval\footnote{In particular, the solution (\ref{A-sol}) is not a good approximation to the vacuum solution $A_+^{\rm vac} = \frac{{\rm e}^{-i k \tau}}{\sqrt{2 k}}$ in the deep sub-horizon regime, where the energy associated to (\ref{A-sol}) vanishes. Therefore, this approximated solution is sometimes employed to effectively renormalize away UV divergencies associated with the energy density of the vacuum modes.} $\left( 8 \xi \right)^{-1} \la k / \left( a H \right) \la 2 \xi $ \cite{Barnaby:2010vf} during which the physical energy density associated to the mode is the largest. We note that $A_{+}$ is exponentially sensitive to $\xi$. Therefore, all the signatures   associated to this production are very sensitive to small variations of $\xi$ in the  relevant $\xi \ga 1$ region.

The original work  \cite{Anber:2006xt} studied under which conditions this production can result in sufficiently large magnetic fields, under the assumption that the gauge field coupled to the axion is    the photon. This issue was further studied in  \cite{Durrer:2010mq}, which reaches   more pessimistic conclusions than  \cite{Anber:2006xt}, based on different assumptions on the inverse cascade post-inflationary evolution of the magnetic field.

A different application was proposed in Ref.  \cite{Anber:2009ua}: the production of gauge fields occurs at the expense of the kinetic energy of the inflaton field, so that, at sufficiently large $\xi$,  the production backreacts on the inflaton field evolution, slowing down the inflaton motion. This can allow for a novel realization   of  the mechanism of warm inflation \cite{Berera:1995ie}. Using a mean field approximation to account for the backreaction of the gauge quanta, the background inflaton evolution is controlled by
\begin{equation}
\ddot{\phi} + 3 H \dot{\phi} + \frac{d V}{d \phi} \simeq \frac{\alpha}{f} \left\langle \vec{E} \cdot \vec{B}  \right\rangle
\end{equation}
where dot refers to derivative with respect to physical time, and here and in the following we continue to use the electromagnetic notation
for the ``electric'' and ``magnetic'' components of the gauge field, even if we no longer assume that the gauge field is our photon (this assumption is clearly necessary for the magnetogenesis application, but it can be dropped for all the other effects that we discuss). While the standard slow roll evolution (without particle production) occurs in the regime of $3 H \dot{\phi}  + \frac{d V}{d \phi} \simeq 0$, it was shown in  \cite{Anber:2009ua} that a sufficiently large $\alpha = {\rm O } \left( 100 \right)$ (to be contrasted with the natural value $\alpha\sim \mathcal{O}(1)$) induces the slow-roll regime  $3 H \dot{\phi} \simeq \frac{\alpha}{f} \left\langle \vec{E} \cdot \vec{B}  \right\rangle$. In particular, slow-roll inflation takes place even if the axion decay constant is subPlanckian and the associated potential $\cos \left( \frac{\phi}{  f} \right)$ is steep. This is hence a small-field model, since $\Delta \phi \sim f < M_{Pl}$.
As we discuss more in details in subsection  \ref{subsec:particle}, this mechanism  is also constrained by  the fact that the produced gauge field also strongly affect the primordial density perturbations. The primordial perturbations have the correct amplitude and are sufficiently close to Gaussian if the rolling axion couples to ${\cal N} = {\rm O } \left( 10^5 \right)$ gauge fields  \cite{Anber:2009ua,Anber:2012du}. 

Ref. \cite{Visinelli:2011jy} also studied    warm inflation with an inflaton axion, but without specifying the  origin of the inflaton decay $\Gamma$, and therefore without including the effects associated to the gauge field nonperturbative production.

 
\subsubsection{Coupling to a non-Abelian gauge field}

Ref. \cite{chromo} showed that the interaction of an axion with an SU(2) gauge field with a non-vanishing vev can also allow inflation for a steep potential even without any gauge quanta production. The coupling with the inflaton results in an effective potential for the gauge vev $Q$; the equation of motion for the inflaton field, with $Q$ settled to the minimum of this effective potential, admits a slow-roll inflationary solution. An analogous modification of the inflaton equation of motion does not take place for the abelian case $g=0$  (where $F_{\mu \nu}^a = \partial_\mu A_\nu^a -   \partial_\nu A_\mu^a   - g \epsilon^{abc} A_\mu^b A_\nu^c$) and for vanishing vev, $Q=0$. Since the mass of the gauge field due to the non-abelian structure  is $m_g \propto g Q$, one may expect that the inflationary solution is unstable for light gauge field.  Ref.  \cite{Dimastrogiovanni:2012st} studied the model in a regime of heavy gauge field, where the latter can be integrated out, and the inflationary solution is stable. A more general study was presented in ref. \cite{DP}, where it was showed that the inflationary solution is stable if and only if $m_g > 2 H$. The instability is an instability of the inflationary solution, and not of the model (the Minkowski solution with $A_\mu^a=\phi=0$ is perfectly stable) and, as a typical instability of a given solution, it manifests itself as a tachyonic growth of one scalar perturbation  \cite{DP}.  Technically, it is due to  
a negative term proportional to $k$ in the dispersion relation of one scalar perturbation, induced by the $\phi F {\tilde F}$ coupling, analogously to the tachyonic growth of the gauge quanta emerging from (\ref{eq-A}). In the current case, however, the instability, when present, takes place inside the horizon and is extremely rapid. A sufficiently high mass for the gauge field shuts off this instability  \cite{DP}.

Ref.  \cite{DP} also noted  that the model can lead to a large production of the  vacuum gravity wave mode, in excess of the standard Lyth bound  $r > 16 \epsilon$, which does not apply in this context, since it holds for a free inflaton and unsourced tensor modes.
The stability result of \cite{DP} was confirmed by \cite{Adshead:2013qp}, that 
also presented a complete study of the gravitational waves produced in the model, showing that they are chiral. This is due to the fact that the axion evolution breaks parity. For the  model \cite{chromo}, the gravity wave production occurs already at the linearized level, due to the mixing with the gauge field perturbations caused by the nonvanishing vector vev. As we review in subsection \ref{ss:GW},  in the abelian case an interesting wave signal, which also violates parity, can instead be sourced at the nonlinear level by the gauge quanta that originates from the $\phi F {\tilde F}$ term.

Ref.~\cite{DP} computed the power spectrum of the scalar perturbations for some illustrative choices of the parameters; for instance, the axion decay constant was fixed to $f= 10^{-2} M_{Pl}$, so that the inflaton potential is indeed too steep to provide inflation. In the regime studied in \cite{DP} it was found that the spectrum becomes too red at the lower values of $m_g / H$ considered. This agrees with the analytical results of  \cite{Dimastrogiovanni:2012st} in the large $m_g / H$ regime. On the other hand,  the gravity wave production increases with $m_g / H$, and can easily exceed the observational bounds  \cite{Adshead:2013qp}. The requests that both the scalar and tensor spectrum are compatible with observations therefore translate in potentially conflicting requests on  $m_g / H$, and, as first realized in  \cite{Dimastrogiovanni:2012st}, this may render the model  \cite{chromo} incompatible with observations. The observation of 
 \cite{Dimastrogiovanni:2012st} has been recently confirmed by \cite{Adshead:2013nka}, that evaluated the perturbation results of 
 \cite{Dimastrogiovanni:2012st,Adshead:2013qp} for an exhaustive scan of parameters, finding that the model of  \cite{chromo} is indeed incompatible with the CMB data.

Refs.~\cite{gauge-flation1} presented a model of inflation with only a non-abelian gauge field, characterized by a $( F {\tilde F})^2$ term in addition to the usual $F^2$ term. At the classical level this model can be understood as the limiting case of \cite{chromo} in the limit in which the axion is heavy and can be integrated out \cite{gauge-flation2}.

 
\subsection{Extradimensional axion inflation}
\label{ss:extradim}

A controlled  effective superPlanckian axion decay constant can also  be obtained starting from a local gauge symmetry in a model with extra dimension(s); Ref.~\cite{ACCR} realized this idea  in the simplest  case of  one extra dimension compactified on a circle of radius $R$. The extra component $A_5$ of an abelian gauge field propagating in the bulk cannot have a local potential, due to  gauge invariance in five dimensions. Therefore the Wilson loop $\theta \equiv  \oint dx_{5} A_5$ cannot receive a local potential contribution in the extra dimensional theory, giving rise to a shift symmetry $\theta \rightarrow \theta + {\rm constant}$.
 The symmetry is however broken non locally through the  so called Hosotani mechanism  \cite{Hosotani:1983xw}  by the Casimir energy of massless fields propagating in the bulk with charge $q$  under the gauge symmetry. This results in the potential  
\begin{equation}
V \left( \theta_c \right)  \propto \frac{1}{R^4} \sum_{n=1}^\infty \frac{ \cos \left( 
2 \pi R g_{4D} q n \theta_c \right) }{n^5 }
\end{equation}
where $\theta_c \equiv \theta / \left( 2\pi g_{4D}R \right)$ is  canonically normalized, and $g_{4D}$ is the four-dimensional gauge coupling. A sufficiently small gauge coupling (in the reduced $4D$ theory) $g_{4D}$ allows for a superPlanckian axion decay constant even if $R > M_{pl}^{-1}$, so that quantum gravity effects are under control \cite{ACCR}. The phenomenology is then the same as that of natural inflation (summarized in figure \ref{fig1}).


\section{Phenomenology}
\label{sec:pheno}

In this section we review some of the phenomenological signatures of models of axion inflation beyond the $\{r,\,n_{s}\}$ plane of figure \ref{fig1}. 
In the first subsection  we study the primordial perturbations and gravity waves originated by   gauge field production during inflation. In the remaining two subsections we discuss the oscillations in the primordial power spectrum and in the higher correlators that  generically arise non-perturbatively in models of axion inflation.


\subsection{Production of gauge quanta and their  inverse decay into inflaton perturbations}
\label{subsec:particle}

The couplings of an axion to other fields are highly constrained by the shift symmetry. The leading operators that control the coupling to gauge and fermion fields are
\begin{equation}
{\cal L}_{\rm int} \supset - \frac{\alpha}{4 f} \phi F_{\mu \nu} {\tilde F}^{\mu \nu} + \frac{C }{f} \partial_\mu \phi   {\bar \psi} \gamma_5 \gamma^\mu \psi
\end{equation}
where $f$ is the axion decay constant, $F_{\mu \nu}$ the gauge field strength (for simplicity, a U(1) gauge field is considered here, although 
several results can be  generalized to the nonabelian case),   ${\tilde F}_{\mu \nu}$ its dual, and where  $C$ and $\alpha$ are  model dependent coefficients that, in the spirit of an effective field theory,  are naturally expected to be ${\rm O } \left( 1 \right)$. The associated perturbative decay rates are
\begin{equation}
\Gamma_{\phi \rightarrow A A } = \frac{\alpha^2 m_\phi^3}{64 \pi f^2} \;\;,\;\;
\Gamma_{\phi \rightarrow \psi {\bar \psi } } \simeq \frac{C^2}{2 \pi f^2} m_\phi m_\psi^2
\end{equation}
The decay into fermions is helicity suppressed, and is typically subdominant (for $m_\psi \ll m_\phi$, as we have assumed) with respect to that into gauge fields. Therefore, one should naturally expect that the reheating after inflation is mostly controlled by the latter process. 

 As we have already discussed in subsection \ref{ss:gauge}, the  production of gauge fields can actually be relevant  already during inflation, and lead to additional friction that can allow for a slow-roll evolution even for $f \ll M_{pl}$    \cite{Anber:2009ua}. The mechanism of  \cite{Anber:2009ua}  is constrained by the requirement that backreaction is relevant and that the primordial perturbations match the amplitude inferred from the CMB temperature anisotropies. The full equation for the inflaton perturbations in the regime of strong backreaction is very hard to solve, and only an approximated form of this equation has been solved in  \cite{Anber:2009ua} and in the subsequent analyses of  \cite{BPP,Anber:2012du,Linde:2012bt}:
\begin{equation}
\delta \ddot{\phi} + 3 \beta H \delta \dot{\phi} - \frac{\nabla^2}{a^2} \delta \phi + \frac{\partial^2 V}{\partial \phi^2} \delta \phi \simeq \frac{ \alpha
}{ f } \, \left[ \vec{E} \cdot \vec{B} - \langle \vec{E} \cdot \vec{B} \rangle \right]
\label{eq-dphi}
\end{equation}
with
\begin{equation}
\beta \equiv 1 - 2 \pi \xi \frac{ \alpha }{ f }  \frac{\langle \vec{E} \cdot \vec{B} \rangle}{3 H \dot{\phi}}
\end{equation}
The right hand side of this equation accounts for the inverse decay of gauge field quanta into inflaton perturbations  \cite{Barnaby:2010vf}, 
which is the only relevant effect in the regime of weak backreaction on the background dynamics. The amount of produced gauge quanta
is given in eq. (\ref{A-sol}).  The second term in $\beta$   accounts for the dependence of $\langle \vec{E} \cdot \vec{B} \rangle$ on $\dot{\phi}$ (via its dependence on $\xi$), and is expected to give the dominant backreaction effect in the evolution equation for $\delta \phi$.  However it remains to be proven whether additional nonlinear terms present in the full system can be relevant in the regime of strong backreaction  \cite{Linde:2012bt}. 

Analogously to the background inflaton, one may expect that, in the strong backreaction regime, also the equations for the perturbations can be solved by only keeping the term $ 3 \beta H \delta \dot{\phi} \sim 3 \beta H^2 \delta \phi$ on the left hand side (with the derivative evaluated at horizon crossing when the inverse decay is strongest) \cite{BPP}. This gives
\begin{equation}
 \langle \zeta^2 \rangle \sim \frac{1}{\left( 2 \pi \xi \right)^2}  \;\;\;,\;\;\; \xi \gg 1
\label{zeta-strong}
\end{equation} 
  for the   curvature perturbation on uniform density hypersurfaces $\zeta$.  This estimate agrees with the precise solution of eq.  (\ref{eq-dphi}) in the  strong backreaction regime \cite{Anber:2009ua,Anber:2012du,Linde:2012bt}. If the axion is coupled to $ {\cal N} $ gauge fields with comparable strength, the above result is modified into  $\langle \zeta^2 \rangle  \sim \frac{1}{\left( 2 \pi \xi \right)^2 \, {\cal N}}$  \cite{Anber:2009ua}.  A large value ${\cal N} = {\rm O } \left( 10^5 \right) $ is assumed in   \cite{Anber:2009ua,Anber:2012du}  to match the observed amplitude of the primordial perturbations.
  

\subsubsection{Inverse decay corrections to primordial correlators}

Refs. \cite{Barnaby:2010vf,Barnaby:2011vw} pointed out that the  gauge field production can result in interesting scalar perturbations already in the regime of weak backreaction. 
The reason is that the inflaton perturbations are sourced by the gauge modes are non scale invariant (as we discuss below) and highly non-Gaussian. This second aspect is due to the fact that they result from the convolution of two almost Gaussian gauge perturbations. The shape of the bispectrum is close to the equilateral one \cite{Barnaby:2011vw}: the cosine (as defined in  \cite{Babich:2004gb}) between the bispectrum from this mechanism and the equilateral template is $\simeq 0.94$. This is due to the fact that, as we mentioned, only modes of size comparable to the horizon are present at any moment during inflation. Therefore, the inverse decay only occurs between modes of roughly equal size, resulting in $\delta \phi$ modes of that size. Modes  $\delta \phi$ of too different wavelength are sourced by different gauge modes, and are therefore uncorrelated. 

In the weak backreaction regime, the inflaton perturbations obey   eq. (\ref{eq-dphi}) with $\beta \simeq 1$. The solution is a superposition of the standard vacuum mode (the solution of the corresponding homogeneous equation) and the mode sourced by $A$. 
The two modes add up incoherently, so that $\langle \zeta^n \rangle$ =  $\langle \zeta^n_{\rm vacuum} \rangle + \langle \zeta^n_{\rm sourced} \rangle $ for the connected $n-$point correlators (Ref. \cite{Barnaby:2011pe} showed that the higher order correlators of the sourced modes scale in a characteristic way with $n$, that may help to distinguish this non-Gaussianity from other models of nearly equilateral non-Gaussianity).

 The fact that the observed primordial perturbations are highly gaussian imposes the requirement that the vacuum modes should dominate over the sourced ones. This happens for  $\xi\la 2.9$  \cite{Barnaby:2011vw},  which corresponds to\footnote{Subsequent analyses have used current observational bounds to constrain other shift symmetric couplings of the inflaton \cite{ABGM}}
 \begin{equation}
\frac{f}{\alpha} \ga {\rm O } \left( 10^{16} {\rm GeV} \right)
\label{bound-f}
\end{equation}
  Given the exponential sensitivity of the gauge field production to $\xi$, see eq. (\ref{A-sol}), all the interesting phenomenological signatures associated to the gauge field production that we discuss here  take place for values of the parameters close to this bound.   Ref. \cite{Meerburg:2012id} computed the limit on $\xi$ resulting from the WMAP7 bound \cite{Komatsu:2010fb} on equilateral non-Gaussianity. Here we update this bound in light of the Planck constraint on equilateral non-Gaussianity $f_{NL}^{\rm eq}=-42\pm 75$. The precise numerical value depends on the inflaton potential, and on the prior expected on $\xi$. Following  \cite{Meerburg:2012id}, we denote by $\xi_*$ the value of $\xi$ when the modes with $k_* =  0.002 \, {\rm Mpc}^{-1}$ left the horizon.  For a  quadratic inflaton potential, a flat prior in the interval $0<\xi_*<10$ results in the bound $\xi_* < 2.45$ for WMAP7 and $\xi_{\ast}<2.37$ for Planck, while a log-flat prior in the interval $10^{-1} < \xi_* < 10^{2}$ results in the bound $\xi_* < 2.22$ for WMAP7 and $\xi_{*}<2.14$ for Planck, all at $95\%$ CL.

Up to constant factors\footnote{We adopt the standard definition of the slow roll parameters $\epsilon \equiv \frac{M_{Pl}^2}{2} \, \left( \frac{1}{V} \, \frac{d V}{d \phi} \right)^2$ and $\eta \equiv M_{Pl}^2 \, \frac{1}{V} \frac{d^2 V}{d \phi^2}$.}, $\xi \propto \frac{\dot{\phi}}{H} \propto \sqrt{\epsilon}$. Therefore, the time evolution of this parameter is  suppressed by slow roll as $\frac{\dot{\xi}}{H \xi} =     2  \epsilon -   \eta  $. 
Despite this suppression,  even a small change of $\xi$ can strongly impact the phenomenology associated to the gauge field production, 
which is exponentially sensitive to this parameter \cite{Cook:2011hg}. In particular, $\epsilon$ typically increases during inflation, so that the gauge field production increases at progressively smaller scales.  For the primordial density perturbations, a first signature of this is an increase of the power of density perturbations  at the smallest observed CMB scales  \cite{BPP}.  To compare with the non-Gaussianity limits, it is useful to refer to the value of $\xi$ at the pivot scale $k_*$, even if this effect is due to the increase of $\xi$ at $k \gg k_*$.  Ref. \cite{Meerburg:2012id} obtained the limits on $\xi_*$ resulting from the WMAP7  \cite{Komatsu:2010fb} and the ACT \cite{Dunkley:2010ge} data.  For a quadratic inflaton potential, the $95\%$ CL limit with flat and log-flat priors (in the same interval as the bounds from non-Gaussianity) are $\xi_* < 2.41$ and  $\xi_* < 2.15$, respectively \cite{Meerburg:2012id}. This limit is therefore slightly stronger than the one obtained from non-Gaussianity. Stronger bounds are expected from the Planck and ACTPol data, where the forecasted  $95\%$ CL limit is $\xi_* < 2.12$ in case of flat prior, and  $\xi_* < 1.92$ in case of log-flat prior   \cite{Meerburg:2012id}.


\subsubsection{Primordial black holes}

A second signature associated with the increase of $\xi$ is the possible formation of primordial black holes \cite{Linde:2012bt}.  This issue was first studied in \cite{Lin:2012gs}, under the incorrect assumption of negligible backreaction of the gauge fields on the background dynamics throughout inflation. The backreaction is negligible for the  values of $\xi_*$ mentioned above, and therefore when the CMB modes leave the horizon.  However,  for typical chaotic inflationary potentials, $\xi$ increases to sufficiently large values so that the latest stages of inflation take place in the regime of strong backreaction \cite{BPP}.  Taking this into account, ref.  \cite{Linde:2012bt} showed
that the primordial black hole limit enforces the bound $\xi_* \la 1.5$.  This limit is much stronger than those from the CMB scales, and also than those from the gravity waves that we discuss below. However, it relies on eq. (\ref{eq-dphi}) being a good representation of the equation for the scalar perturbations in the strong backreaction regime.

\begin{figure}
\centering
\includegraphics[width=0.5\textwidth]{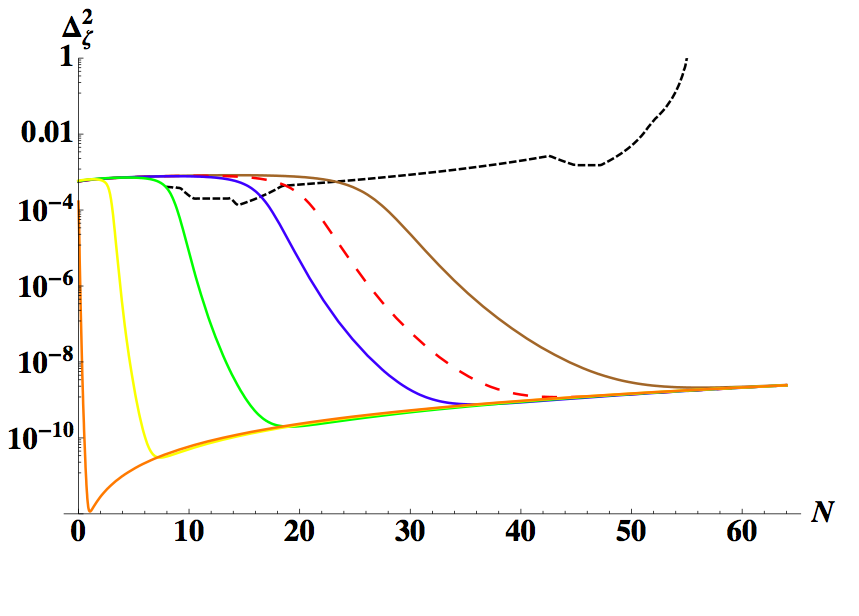} 
\includegraphics[width=0.48\textwidth]{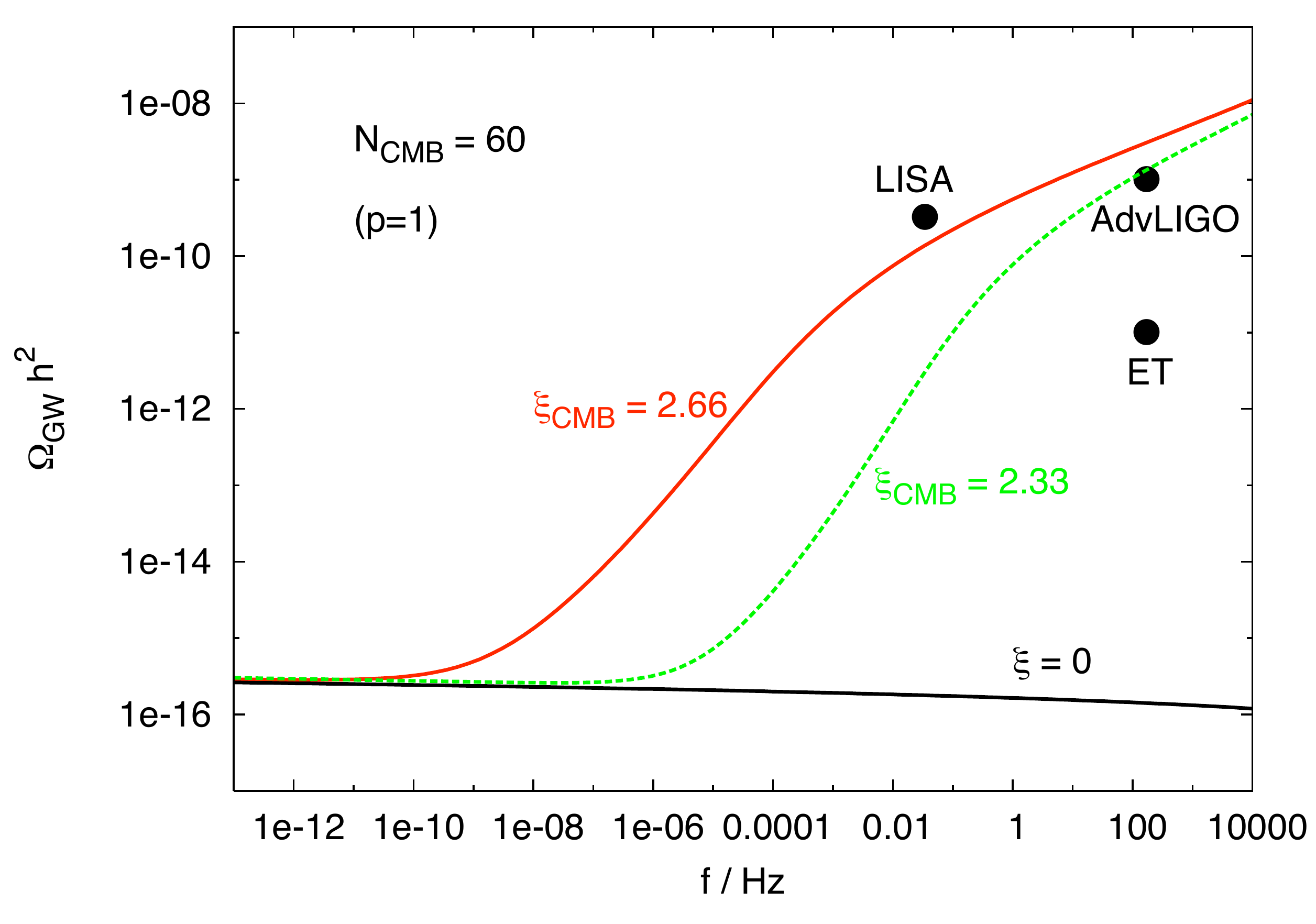}
\caption{The left plot, taken from \cite{Linde:2012bt}, shows the primordial scalar power spectrum as function of efoldings before the end of inflation including the corrections coming from the inverse decay of gauge fields together with the upper bound coming from the constraints on primordial black holes. 
We note a growth of the power spectrum with decreasing scales (corresponding to decreasing $N$) from the COBE normalized value to the saturation value (\ref{zeta-strong}) in the strong backreaction regime. 
The right plot, taken from \cite{BPP}, shows the fraction of the total energy density in terms of primordial gravitational waves as function of frequency produced by the inverse decay of gauge fields for the three indicated values of $\xi$ and a linear potential. Also in this case we signal grows at decreasing scales (corresponding to higher frequencies.  For comparison we also show the expected sensitivity of three gravitational wave experiments (see \cite{BPP} for details).  \label{fig2}}
\end{figure}

 
\subsubsection{Primordial gravitational waves}

\label{ss:GW}

The gauge quanta produced by the inflaton through the $\phi F {\tilde F}$ term source also tensor modes
 \cite{Barnaby:2010vf}. The sourced signal is parity violating: one chirality is produced with a much greater abundance than the other one
 \cite{Sorbo:2011rz}. In principle, this parity violation may be detectable in CMB experiments through  non-vanishing TB and EB correlators 
\cite{Saito:2007kt,Gluscevic:2010vv}, provided that $r$ is sufficiently large and that the sourced gravity waves constitute a sizable portion 
of the total gravity waves  from inflation. In the simplest implementation of the mechanism, this is incompatible with the limits on $\xi_*$ mentioned above from the scalar perturbations at the CMB scales. Such limits can be evaded if the primordial perturbations are sourced by a curvaton field 
 \cite{Sorbo:2011rz}, or if the axion that generates the gauge quanta is not the inflaton \cite{Barnaby:2012xt}. In this way, one minimizes the amount of sourced scalar density perturbations by avoiding a direct coupling between them and the gauge quanta. Still, a gravitational coupling cannot be avoided, but ref.  \cite{Barnaby:2012xt} verified that the scalar  perturbations generated gravitationally can be kept at a sufficiently small level, while the sourced gravity waves can be significant.

Ref.  \cite{Cook:2011hg} pointed out that, identically to what happens for the scalar perturbations, also the gravity wave signal sourced by the gauge quanta grows at smaller scales, as the production of the gauge quanta grows during inflation. The sourced gravity waves may be  detectable at terrestrial interferometers. A precise computation of this signal requires taking into account the fact that, for values of $\xi$ that lead to an interesting amount of gravity waves, the final stages of inflation occur in the  strong backreaction regime \cite{BPP}.  By taking this into account, ref.  \cite{BPP} estimated that values of couplings compatible with the CMB limits from the scalar modes may lead to an observable signal already at Advanced LIGO. A more detailed study was performed in ref.   \cite{Crowder:2012ik}. It was shown there that, for a quadratic inflaton potential,  the forthcoming second generation gravity waves experiments (Advanced LIGO, VIRGO, and Kagra)   can probe $\xi_* \ga 2.2$ (at $95 \% $ CL, based on flat priors on $\xi$), while third generation experiments can improve the limit to   $\xi_* \ga 1.9$. These forecasted limits are better than the current CMB limits of  \cite{Meerburg:2012id}, and, for the third generation, better than the forecasted CMB limits based on Planck and ACTPol. The gravity waves limits are weaker than the constraint from the primoridal black holes  \cite{Linde:2012bt}, although they are free from the uncertainty associated to the latter.


\subsection{Resonant oscillations in all primordial N-point functions}
\label{ss:res}

As pointed out in section \ref{sec:models}, the shift-symmetry is broken non-perturbatively so that one expects non-perturbative corrections to the axion potential in the form of a sinusoidal modulation. Despite being typically a small correction to the background dynamics, these terms can imprint a very characteristic oscillating signal in all primordial correlators. The signal, induced by a resonance between background and perturbations \cite{FMPWX,FP}, has three parameters: the amplitude related to the size of non-perturbative effects; the frequency in $\ln (k/k_{\ast})$, which is given by $\alpha\equiv \omega/H=\sqrt{2\epsilon}M_{Pl}/f$ with $\omega$ the time frequency of background oscillations and $f$ the axion decay constant; the phase of the oscillation, which is arbitrary and is always marginalized over in comparing with observations. Since perturbations freeze outside of the horizon, in order to have a resonance one needs $\omega>H$ and hence $\alpha>1$. The resonance then takes place at $-k\tau=\alpha/2$, which is well inside of the horizon when $\alpha\gg 1$ and the signal is stronger. We separate our discussion of the observational constraints by discussing first the power spectrum and then the bispectrum.

 
\subsection{Oscillations in the power spectrum}

After \cite{CEL} recognized the observational importance of background oscillations (with emphasis on the non-Gaussian aspects to be discussed in the next subsection), in \cite{FMPWX} it was argued that these oscillations generically arise non-perturbatively in models of axion inflation (see also \cite{sur} for an earlier discussion) and their effect on the power spectrum was computed analytically showing that the underlying mechanism is a resonance between the perturbations and the background. The signal was searched in WMAP5 data finding two candidate frequencies with $\Delta \chi^{2}\sim 10$ \cite{FMPWX} (see also \cite{KLP} for other constraints on oscillations in the power spectrum). Subsequently $\Delta \chi^{2}\simeq 13$ was found for WMAP7 \cite{AHSS}. An update with WMAP9 data shows shows again a preference for the same two frequencies \cite{EFP}, with $\ln (f/M_{Pl})\simeq -3.38$ leading to $\Delta \chi^{2}\simeq 19$ when the normalization of the primordial power spectrum is allowed to vary (as it should). It would be interesting to estimate the significance of this improvement using simulated data. At the time of writing, it seems that such a large improvement is \textit{not} present for Planck data \cite{EF}. Note that real space analysis of CMB data does not lead to any advantage in extracting these oscillations because of the logarithmic spacing in momentum space \cite{BJW}. Forecast for constraining oscillation in the power spectrum with large scale structure observables were performed in \cite{HVV}.

 
\subsection{Oscillations in higher $N$-point functions}

As we mentioned, the authors of \cite{CEL} first showed numerically that background oscillations during inflation can lead to large non-Gaussianity of a particular modulated shape. An analytical calculation from first principles of this \textit{resonant non-Gaussianity} was provided in \cite{FP} (see also \cite{HHJS}), where it was shown that it is orthogonal to any other smooth non-Gaussian shape (the cosine defined in \cite{Babich:2004gb} being bounded from above by $\alpha^{-1}\ll 1$). These results were later generalized to all $N$-point function up to remarkably large $N\sim 20$ \cite{LP}. The consequences of oscillating backgrounds were also discussed in the context of the Effective Field Theory of inflation in \cite{BDMS}, where it was pointed out that at least in momentum space (as opposed to multiple space) the signal to noise ratio of resonant non-Gaussianity is bounded from above by the signal to noise ratio of the (correlated) oscillations in the power spectrum. One can hence look for resonant non-Gaussianity using the same two sets of parameters (amplitude, frequency and phase) that lead to the improved fit in the power spectrum discussed in the previous section, corresponding to $f_{res}=700$ and $f_{res}=250$ for the higher and lower frequency, respectively. The constraints on $f_{res}$ from the CMB temperature bispectrum have not yet been derived (but see \cite{Daan} for progress in this direction) in this interesting range of high frequencies, but they can provide a precious check to confirm or reject the hints coming from the power spectrum. A forecast of the expected constraints on the resonant non-Gaussianity from the scale-dependent halo bias was given in \cite{CS}. Some phenomenological properties of many superimposed resonant non-Gaussianities have been considered in \cite{GRW}.

 
\section{Discussion and outlook}
\label{sec:discussion}

A prolonged phase of primordial inflation that can explain the current data requires small slow-roll parameters. This means that the ``mass'' of the inflaton has to be much smaller than its natural scale, namely the Hubble scale $H$ during inflation. To address this problem, often referred to as the  \textit{$\eta$-problem}, one can either rely on fine tuning of the potential or invoke a symmetry. When the inflaton displacement is larger than the Planck scale, namely in large-field inflation, infinitely many higher dimensional corrections are expected on general grounds and hence the required fine tuning is infinite. For large-field models, which are phenomenologically distinguishable because they lead to observable primordial tensor modes, the use of symmetry is the only viable possibility. It is well known that supersymmetry is not sufficient since it is broken at the Hubble scale by the deSitter background. Hence we are lead to consider models with a shift symmetry. In this short review we have discussed several proposals for the theoretical framework from which a shift symmetry can arise and how predictions compare with current data. Many scenarios are compatible with current observations at $95\%$ CL or better, including models with more than one axion, models with non-minimal dynamics and models with controlled breaking of the shift symmetry. 

In the near future, CMB polarization experiments will improve the sensitivity to primordial tensor modes by at least an order of magnitude. A detection will provide strong support in favor of large-field models hence substantiating the need for a shift symmetry. On the other hand, it is natural to ask if a shift symmetry can still play a role in small-field models, which will be favored if we will reach an upper bound $r\lesssim 0.01$. Although a shift symmetry certainly helps in controlling corrections to small-field models, it is not by itself sufficient to avoid fine tuning at the order of a part in a hundred. This can be understood in different ways. As opposed to large-field models, which work with monomial potentials, small-field models rely on the precise cancellation of the slope of two or more leading terms to flatten the potential in a small interval of field space. While the shift symmetry can suppress additional corrections to these leading terms, it does not help in tuning their relative size. A different argument goes as follows. If the shift symmetry arises below some scale $f$, as in the case of a Nambu-Goldstone boson, the typical curvature of the potential is $ \propto f^{-1}$. In order to get small-field slow-roll inflation one needs $\eta\ll1$ around some approximate critical point of the potential (e.g.~a maximum or a flat inflection point) implying $f\gg M_{Pl}$. But this would lead to a prolonged phase of chaotic inflation near the minimum that would govern the predictions from the last 60 efoldings of inflation. Therefore to get a genuine small-field inflation model one needs $f<M_{Pl}$ and some additional symmetry that suppresses the mass scale well below $f^{-1}$, e.g.~around some local maximum. This problem is equivalent to the little hierarchy problem for the Higgs mass \cite{Barbieri:2000gf}, so the same mechanisms  can be used, such as the little Higgs construction as discussed in \cite{ACCR2,KW}. One way or another, the observations in the next few years will greatly help us narrow down the set of possible mechanism at work in the dynamics of the early universe.

 
\section*{Acknowledgments}

We are thankful to P.~Adshead, R.~Flauger, E.~Silverstein, and L.~Sorbo for useful discussions and to D.~Baumann for the detailed comments on the manuscript. E.~P.~is thankful to the KITP at UCSB (NSF Grant No. NSF PHY11-25915) and to the organizer of the Primordial Cosmology workshop for the hospitality while this review was completed. E.~P.~is supported in part by the Department of Energy grant DE-FG02-91ER-40671.    MP is supported in part by DOE grant DE-FG02-94ER-40823 at the University of Minnesota.   MP would like to thank the University of Padova,  INFN, Sezione di Padova, and the Cosmology Group at the Department of Theoretical  Physics of the University of Gevena for their friendly hospitality and for partial support during his sabbatical leave.


\end{document}